\documentclass{sig-alternate-2013}

\newfont{\mycrnotice}{ptmr8t at 7pt}
\newfont{\myconfname}{ptmri8t at 7pt}
\let\crnotice\mycrnotice%
\let\confname\myconfname%

\permission{Permission to make digital or hard copies of all or part of this work for personal or classroom use is granted without fee provided that copies are not made or distributed for profit or commercial advantage and that copies bear this notice and the full citation on the first page. Copyrights for components of this work owned by others than ACM must be honored. Abstracting with credit is permitted. To copy otherwise, or republish, to post on servers or to redistribute to lists, requires prior specific permission and/or a fee. Request permissions from Permissions@acm.org.}
\conferenceinfo{MM'15,}{October 26--30, 2015, Brisbane, Australia.} 
\copyrightetc{\copyright~2015 ACM. ISBN \the\acmcopyr}
\crdata{978-1-4503-3459-4/15/10\ ...\$15.00.\\
DOI: http://dx.doi.org/10.1145/2733373.2806375}

\graphicspath{{eps/}}

\usepackage{paralist}

\usepackage{subcaption}

\usepackage{multirow}

\newcommand{\etal}{{\em et al. }}

\newcommand{\argmax}{\operatornamewithlimits{\mathit{argmax}}}

\usepackage{color}

\usepackage[pagebackref=false,colorlinks,hidelinks,bookmarks=false]{hyperref}

\begin{document}
%

\title{Pinterest Board Recommendation for Twitter Users}
%
%
%
%
%

\numberofauthors{3} 
%
\author{
%
%
\alignauthor Xitong Yang\\
\affaddr{University of Rochester, Rochester, NY 14623}
\email{xyang35@cs.rochester.edu}
\alignauthor Yuncheng Li\\
\affaddr{University of Rochester, Rochester, NY 14623}
\email{yli@cs.rochester.edu}
\alignauthor Jiebo Luo\\
\affaddr{University of Rochester, Rochester, NY 14623}
\email{jluo@cs.rochester.edu}
}


\maketitle
\begin{abstract}
    Pinboard on Pinterest is an emerging media to engage online social media users, on which users post online images for specific topics. Regardless of its significance, there is little previous work specifically to facilitate information discovery based on pinboards. This paper proposes a novel pinboard recommendation system for Twitter users. In order to associate contents from the two social media platforms, we propose to use MultiLabel classification to map Twitter user followees to pinboard topics and visual diversification to recommend pinboards given user interested topics. A preliminary experiment on a dataset with 2000 users validated our proposed system.
\end{abstract}

\category{H.3.5}{Online Information Services}{We-based services}

\terms{Experimentation}


\section{Introduction}

Socially Curation Service \cite{Kimura:2013:ICD:2502081.2502149}, on which people can share and reshare online content that they found interesting, is becoming more and more popular. For example, Pinterest has successfully attracted more than 40 million monthly active users \footnote{eMarketer, Feb 2015} to create pinboards and share online images. Social curation creates high quality contents for users to discover, search and follow what they are interested in. For example, Pinterest has already delivered better search results than Google for a number of segments, such as food recipes, fashion and DIY \footnote{\url{http://goo.gl/PCWG3W}}. Pinboard, on which people can put together online images for specific topics to share with others, is the key element for Pinterest to engage users. Different from traditional photo albums, such as those on Facebook and Flickr, pinboards are not yet another way for people to organize images, but also encouraging users to carefully select images using their creativity. Also, in order to continuously attract followers, pinboard curators take efforts to keep up most recent contents, so pinboards give users a new way to follow most fresh updates on the topic they are interested in. As with other online systems, efficient information discovery is one of the major hurdles for continuous growth. In this paper, we propose a pinboard recommendation system to help users to discover interesting pinboards on Pinterest.

\begin{figure}[t]
    \centering
    \includegraphics[width=.99\columnwidth]{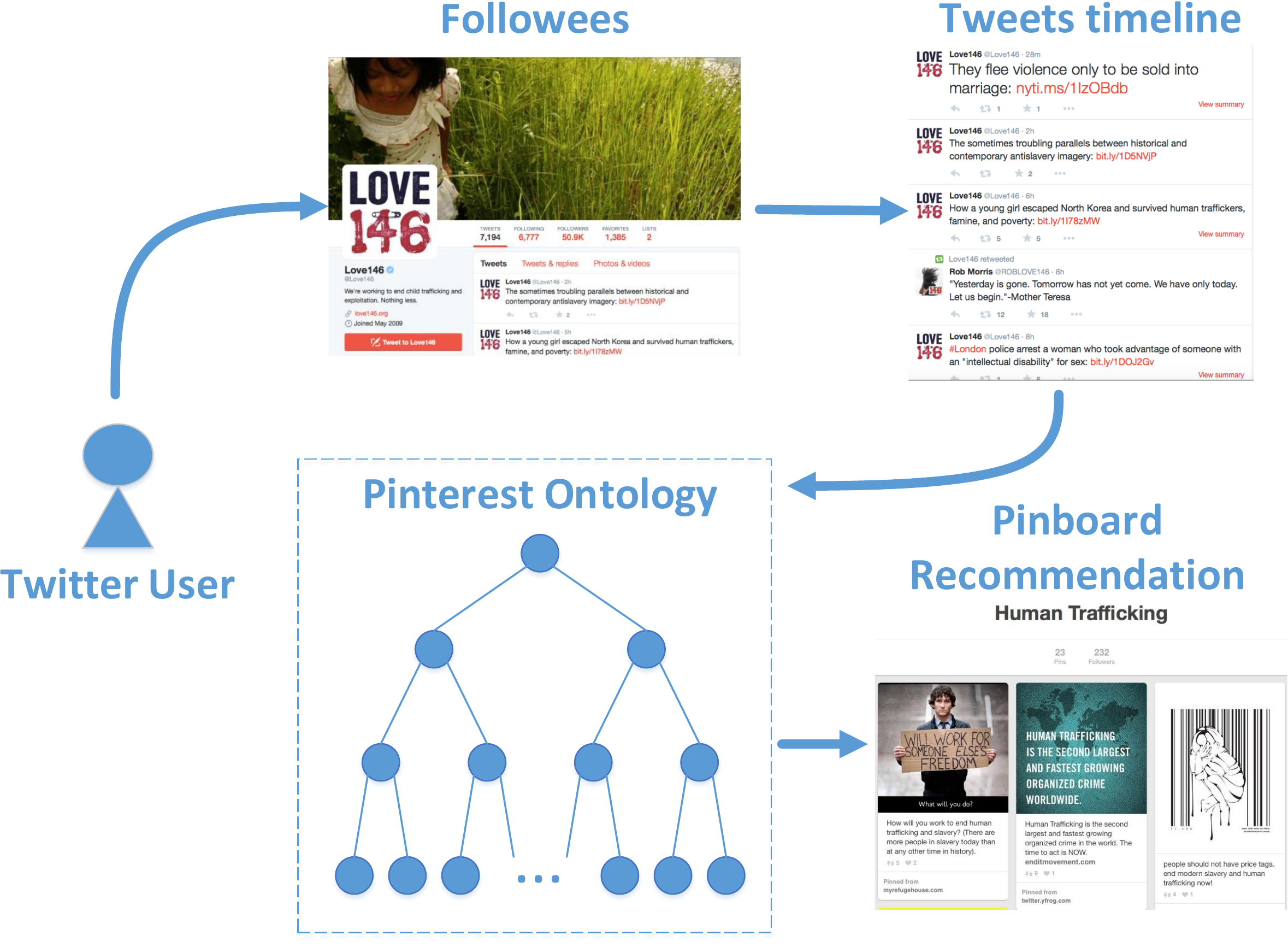}
    \caption{An illustrative example to recommend pinboards to a Twitter user ``@xyz''}
    \label{fig:example}
\end{figure}

\begin{figure*}[t!]
    \centering
    \includegraphics[width=.95\textwidth, height=.25\textheight]{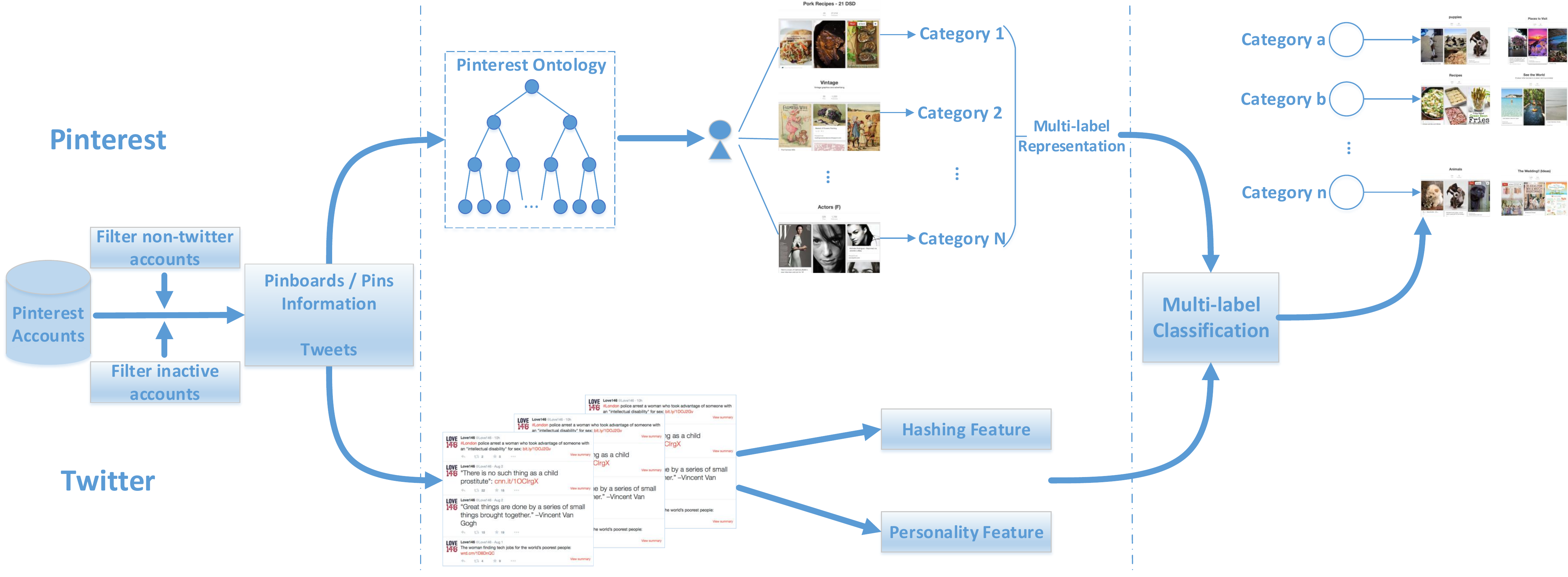}
    \caption{Overview of the proposed recommendation system. First, we align and filter the data from two different sources. After that, we collect multimedia contents, such as Pinboards and Pins information, and tweets from these active users. In the meanwhile, we build the pinboard topic ontology automatically, which is done by pruning an expert ontology such as Wikipedia. Second, we perform user modeling in two folds. For Pinterest, we obtain multi-label representation for each account by mapping their boards to the topics in Pinterest ontology. For Twitter, we extract both text feature and personality feature from user timeline. Third, we associate these two user models using multi-label classification, in which Twitter feature is used to predict the Pinterest category representation of each user. In the end, we recommend diverse Pinterest boards according to this prediction, where diversity is guaranteed by combining visual contents of the images in each board.}
    \label{fig:overview}
\end{figure*}

As a relatively new social media platform, Pinterest is still in its young age. Therefore, how to keep new users is an important topic, but recommendation for new users suffers cold start problem for the lack of activity history. We propose to leverage 3rd party social media platforms, such as Twitter, to solve the cold start problem for new Pinterest users. There are a number of reasons to use Twitter as an intermediate platform to recommend pinboards. First, Twitter is more mature and attracts even more active users, so it represents larger coverage for online users. Second, because most contents on Twitter are text based, it is much easier to find out interesting topics on Twitter than on Pinterest. Third, even though text based platform makes it easier to find interesting topics, image based pinboards, if well aligned with the interests, provide better content for users to digest. For these reasons, recommending pinboards based on Twitter activity history is a perfect match.

We propose following procedure to recommend pinboards to Twitter users.
\begin{inparaenum}[(1)]
\item Given Twitter users, we extract their followees. We assume that the follower relationship represents user interests.
\item Using the proposed association method, we map the followees to pinboards topics, which represent what kind of boards the Twitter users will be interested in. The noise of individual followee can be reduced by aggregating all mapping results.
\item We choose a number of pinboards from the selected topics, and recommend these pinboards to the Twitter users.
\end{inparaenum}

Using the example in Fig.~\ref{fig:example}, we find that twitter user ``@xyz'' is following a twitter user ``@love146''. By collecting and classifying timeline of ``@love146'', we map the user ``@love146'' to one of the pinboard topics ``anti human trafficking'', so we assume that the target twitter user ``@xyz'' is interested in the boards about ``anti human trafficking''. Therefore, we select a number boards on this topic. We further improve the recommendation quality by a reranking method based on visual diversity analysis.

Since the followees can be extracted trivially from Twitter API, we solve the following two key problems in this paper, as illustrated in Fig.~\ref{fig:overview}

\noindent \textbf{Map user timeline to pinboard topics} is an example of cross network analysis \cite{Yan:2014:MCA:2647868.2654920}, and we build the association by mining users who are active on both Pinterest and Twitter. We crawl the timeline and pinboards associated with users that are active on both platforms. The pinboards are then mapped onto a predefined topic ontology using text based matching. After that, the users are associated with a twitter timeline and a distribution of pinboard topics. Then we learn a multi-label classifier to map user timeline to the pinboard topics. Using this multi-label classifier, we can map twitter followees without Pinterest account to pinboard topics.

\noindent \textbf{Pinboard reranking} is the process to recommend the best subset from the list of interested pinboards. We adapt a clustering based algorithm to perform this reranking, and the goal is to make sure that the recommended pinboards cover as much aspects as possible for the selected topics.

We make following contributions in this paper,
\begin{inparaenum}[(1)]
\item We present and release a linked dataset that associate pinboards and Twitter users. This will be of interests to broader research than cross network recommendation.
\item We propose a cross network recommendation method based on multi-label classification.
\item We propose a visual reranking method to diversify pinboard recommendation.
\end{inparaenum}

\section{Related Work}
Our work is related to cross network analysis, user modeling and hierarchical multi-label classification. In this section, we highlight the novelty of our work by comparing with recent works on these related fields.

\noindent \textbf{Cross network analysis} aims to merge social signals from different network to increase online social media platform engagement. For example, in \cite{Yan:2014:MCA:2647868.2654920}, Yan \etal proposed to identify the best Twitter accounts to promote YouTube videos, by mining the associations between topics learning from user tweets and their favorite YouTube videos.

\noindent \textbf{User modeling} is the foundation of personalized services, such as personalized recommendation, search engine reranking and advertisements targeting. One component of our Pinterest board recommendation system is based on Pinterest board ontology mapping, which is inspired from \cite{Geng:2014:OKU:2647868.2654950}. In this work, Geng \etal proposed a multi-task CNN to map pinterest images to fashion ontology and the classification results were sorted as user profiles for image recommendation. We adapt the idea that user interests can be represented by multiple nodes on ontology and further extend the ontology domain to 20 categories, including ``Fashion'', ``Food'' and ``Wedding''. In addition, instead of single label classification on images and aggregating the classification results to form user profile, we perform hierarchical multi-label classification on aggregated user tweets to figure out the user interests directly.

\noindent \textbf{Hierarchical MultiLabel Classification} is the problem of classifying data instances to multiple labels or attributes, in which the labels are structured in a hierarchical taxonomy. \cite{Ren:2014:HMC:2600428.2609595} proposed a hierarchical multi-label system to classify short texts (e.g., tweets). In this work, Ren \etal proposed to use text expansion, e.g., entity linking, to deal with the shortness and concept drift problem in short text classification. Our twitter user modeling is also based on hierarchical multi-label classification. However, instead of single tweet classification, we deal with the entire timeline of users, so that each user is modeled by a large number of tweets. In this paper, we adapt Randomized Labelsets \cite{9783540749578} to efficiently model the hierarchical dependency automatically.

\section{Pinboard Visual Diversification}
The goal of this stage is to recommend most relevant and diverse boards, given predicted categories and millions of board candidates sampled from Pinterest board pool. To guarantee the relevance, we map each board to the categories onto the topic ontology with the method described in Experiments Section~\ref{sec:exp}, and also sort the boards of each category according to their popularity (e.g. followers). We then chose top $k$ boards for each category as its preliminary candidates (we set $k=10$ in our experiment).

To enhance diversity, we utilize the visual contents in Pinterest, that is, for each category, we cluster all pin images that belong to its preliminary board candidates, and then chose the boards that can maximize the coverage of these clusters. Specifically, we use deep Convolutional Neural Network (CNN) to learn useful features for the pin images (fc7 layer of the AlexNet \cite{krizhevsky2012imagenet}). After that, Affinity Propagation algorithm \cite{frey07affinitypropagation} is applied to cluster 4096D image features. After obtaining the clusters and assignment for each pin image, we can build a cluster distribution for each board. Then the board selection problem can be formalize as follows.

Denote the candidate set by $\mathcal{B}$, its subset by $B \subset \mathcal{B}$,  a board in the subset by $b \in B$ and the cluster assignment distribution of the board by $\mathcal{P}_b \in R^C$. $C$ is the number of clusters for all images in the pinboard candidates. Then, we define the set level entropy by,
\begin{equation}
    T(B) = H\left(\frac{\sum_{b \in B} \mathcal{P}_b}{|B|}\right),
\end{equation}
in which $H(x)$ is the entropy of a distribution $x$. Then, we select the best set of pinboards by $\argmax_B T(B)$. We can prove that this is actually an NP-complete problem, so an approximation algorithm is needed, but in our experiments we just use brutal force to enumerate all possibilities by restricting the number of pinboard candidates $|\mathcal{B}|$ and the recommendation set $|B|$.

\begin{table}
\setlength{\abovecaptionskip}{-0.001cm} 
     \setlength{\belowcaptionskip}{-0.6cm} 
    \centering
    \begin{tabular}{l | c}
        \hline
        \# & statistics \\
        \hline
        total nodes & 471 \\
        root nodes & 20  \\
        leaf nodes  & 440 \\
        least/largest depth & 2 / 5 \\
        \hline
    \end{tabular}
    \caption{Pinterest topic ontology Statistics}
    \label{tab:node-stats}
\end{table}

\section{Experiments}
\label{sec:exp}

\subsection{Data Collection}

\subsubsection{Pinterest ontology construction and refining}

Ontology is a natural way to model the hierarchical structure of the categories in Pinterest, as is also used in \cite{Geng:2014:OKU:2647868.2654950} to organize items in fashion domain. To construct a full Pinterest ontology, we first build a preliminary ontology based on Wikipedia Categories, where the root nodes of it are 20 manually selected categories according to the 38 original categories given by Pinterest, such as ``Fashion'', ``Food'' and ``Wedding''. After that, we prune the ontology to adapt it to Pinterest user interest distribution. Basically, categories with low term frequency in board and pin information are removed. As board information and pin information share different weights for the ontology, we consider their pruning approaches separately, and the pruned ontology is obtained by uniting the results of two approaches. In addition, popularity metrics like the number of followers, likes, comments and repins are also considered as positive weighted metric while pruning. As a result, we obtain the refined Pinterest ontology and the statistics are shown in Table \ref{tab:node-stats}.

\subsubsection{Map pinboards to topic ontology}
With the refined Pinterest ontology, we can model a user's Pinterest profile by topic distribution. As users have their own boards, we can obtain this representation by mapping their boards to the categories in the ontology. For each board, this mapping approach takes three steps. First, we match board information to the categories, which can get a few or even no matches (but more accurate), as the board information is always rare. Second, we match all pins information under the board. Third, concatenate the result of previous two steps. After obtaining the categories mapping for each board, we can get the representation for the users through the union of the categories of their own boards.

\subsubsection{Twitter timeline features}

We extract two kinds of features to represent the tweets information for each user. First is the original text feature extraction, where feature hashing \cite{Weinberger:2009:FHL:1553374.1553516} and tf-idf term weighting are applied to reduce feature dimension and weight feature appropriately. Second, we also compute 64 LIWC features (e.g., word categories such as ``social'' and ``work'') to model the personality of each user \cite{pennebaker2001linguistic}. These features may potentially reflect the distribution of their interest and user-curated data (e.g., Pinterest boards).

\subsubsection{Data Statistics and Thresholds}

In data preparation stage, we filter out those accounts that do not have twitter links or twitter activities are not active enough (number of tweets less than 200). Finally we get 2265 accounts from original 50000 sample Pinterest accounts. With these active accounts, we crawl 4.9 million tweets, 25.5 thousand boards and 2 million pins in total.

In ontology pruning stage, we set the threshold of term frequency for pins information to be 200 and the threshold for board information to be 1/100 of it, as the number of boards is nearly 1/100 of that of pins. After that we get 471 nodes in the ontology, as illustrated in Table \ref{tab:node-stats}.

In user modeling stage, for Pinterest data, we build up a multi-label representation for each user. Table \ref{tab:data-stats} shows certain standard multi-label statistics of these representation, such as the number of labels, the label cardinality and the label density \cite{9783540749578}. Label cardinality (LC) is the average number of labels per example, and label density is LC divided by the number of labels $|L|$. For Twitter data, we extract bag of words (BOW) text feature, and then apply feature hashing to get 5000 dimensional features. 64D LIWC feature is also extracted from the tweets via LIWC dictionary.

\begin{table}
    \centering
    \resizebox{8cm}{!}{
        \begin{tabular}{c c c c c}
            \hline
            Examples & Attributes & Labels & Label Cardinality & Label Density \\
            \hline
            2265 & 5000+64 & 471 & 16.02 & 0.034 \\
            \hline
        \end{tabular}
    }
    \caption{Data Statistics}
    \label{tab:data-stats}
\end{table}

\subsection{Preliminary Evaluation}

We use \textit{macro-F1}, a macro-averaged version of F-measure to measure our approaches, which is widely used in multi-label classification tasks\cite{read2010scalable}. Formally, we represent an association of labels $v$ and its prediction $\hat{v}$ as two binary vectors $\{0,1\}^T$. Then F-measure (F1) can be used to measure the accuracy of the prediction, which is the harmonic mean of precision and recall and defined as follows: $F1(v, \hat{v}) = (2\times precision \times recall)/(precision+recall)$, where \textit{precision} is defined as $|v\wedge\hat{v}|/|\hat{v}|$, and \textit{recall} is $|v\wedge\hat{v}|/|v|$.

For a multi-label classification task with $N$ examples and $L$ labels, we have two different ways to average the F1 scores:  
\begin{inparaenum}[(1)]
\item \textit{F1 macro-ex.} average F1 scores of N examples, which reflects example based evaluation.
\item \textit{F1 macro-label:} average F1 scores of L labels, which reflects label based evaluation.
\end{inparaenum}



We compare two common multi-label classification methods: \textit{Binary Relevance}(BR) and \textit{Label Powerset}(LP)\cite{read2010scalable}. BR considers the prediction of each label as an independent binary classification task, while LP considers subset of labels as a single label to perform a multi-class classification task. For efficiency consideration, we adopt RAkEL\cite{9783540749578} method, which is a variant of LP that ensemble different light LP classifiers trained by small random subsets of the set of labels.

In order to evaluate the effect of personality features, i.e., LIWC feature, in this user-interest relevant task, we compare following three cases: \textit{Hashing feature only}, \textit{LIWC feature only} and an \textit{early fusion} of both features.

\subsection{Experiment results and Conclusion}

\begin{figure}[t]

    \centering
    \begin{subfigure}[b]{.45\columnwidth}
        \includegraphics[width=.95\textwidth]{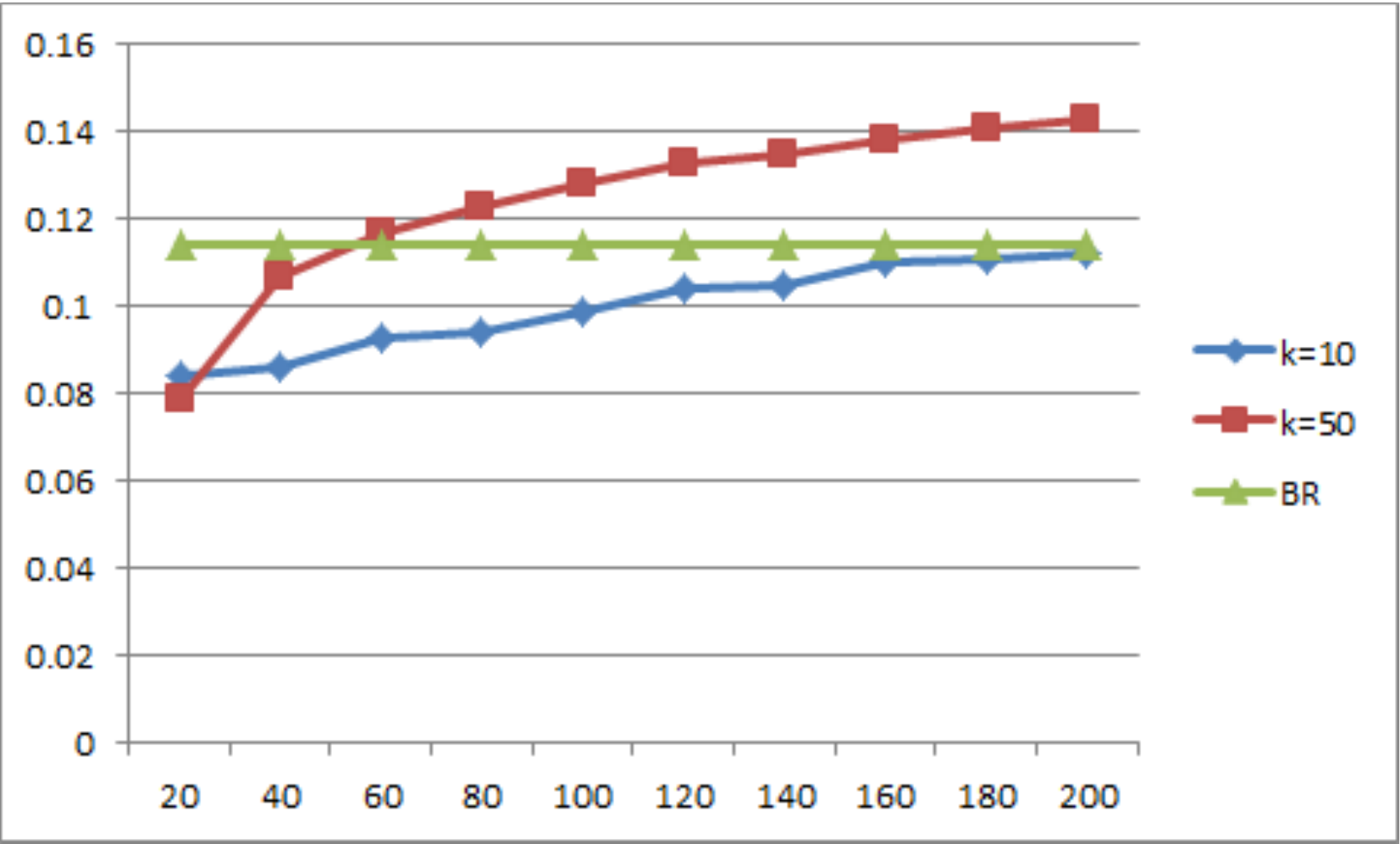}
        \caption{F macro-ex.}
        \label{fig:res-ex}
    \end{subfigure}
    ~
    \begin{subfigure}[b]{.45\columnwidth}
        \includegraphics[width=.95\textwidth]{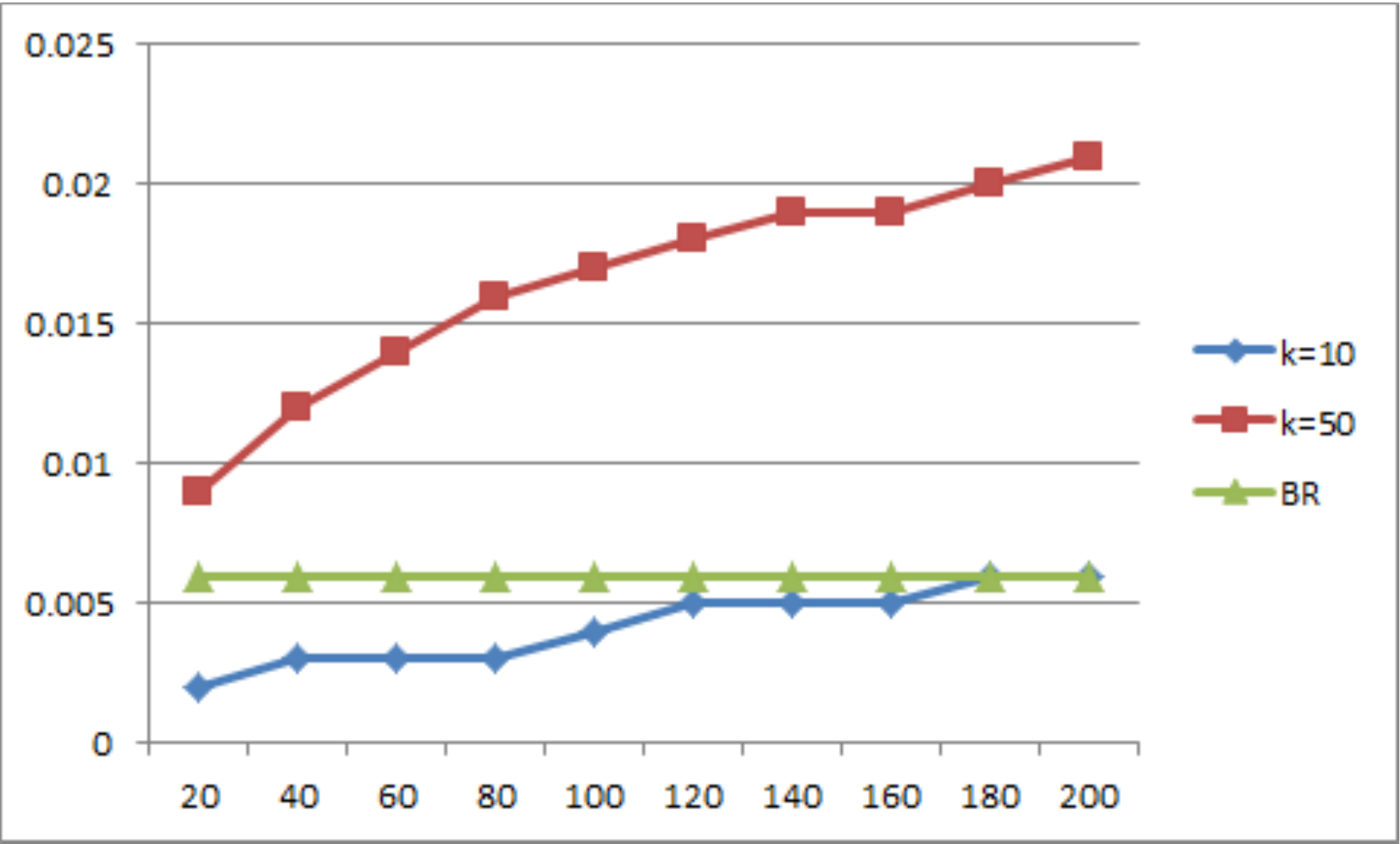}
        \caption{F macro-label}
        \label{fig:res-l}
    \end{subfigure}
    \caption{Experiment results varying key parameters in RAkEL algorithm \cite{9783540749578}.}
        \label{fig:result-tune}
\setlength{\belowcaptionskip}{-0.5cm} 
\end{figure}

Figure.~\ref{fig:result-tune} present the comparison between BR and RAkEL while tuning the parameters k and M in RAkEL method.

\begin{table}
\centering
\begin{tabular}{r|cc|cc}
    & \multicolumn{2}{|c|}{\textbf{F1 macro-ex.}} & \multicolumn{2}{|c}{\textbf{F1 macro-label}} \\
 & BR&LP & BR&LP\\
\hline
Hashing&0.114&0.143& 0.006 &  0.021 \\
LIWC&0.099&0.086& 0.002 & 0.002  \\
fusion&0.115&0.142& 0.006  &0.021  \\
\end{tabular}
\caption{Experiment results, comparing different feature schemes. Note that the total number of labels is 471, so these results are significantly better than random guess.}
\label{tab:results}
\end{table}

Table \ref{tab:results} shows the full experiment results while comparing BR and LP using different kind of features. Through result above, we notice that LP based method can capture label dependencies and get better performance than BR method, especially when the subset size is large enough to contain sufficient labels. LIWC alone cannot get good result, which means that this personality feature cannot reflect one's interest distribution across different platforms.

\begin{figure}[t]
    \centering
    \includegraphics[width=.99\columnwidth]{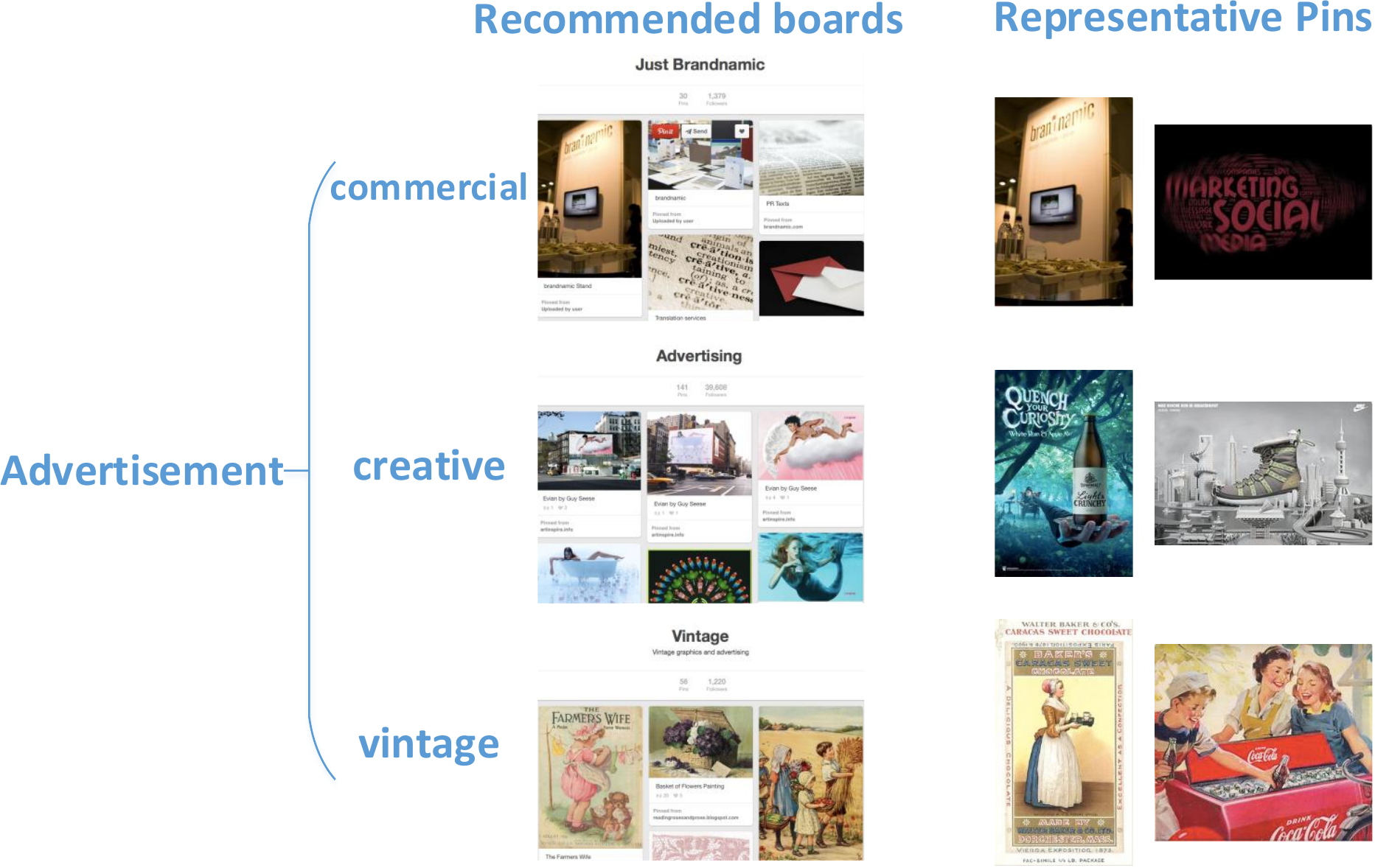}
    \caption{The recommended boards for the topic ``Advertisement''. Different kinds of advertisements are recommended, such as ``commercial Ads'', ``creative'' and ``vintage''.}
    \label{fig:sort}
\end{figure}

We show diversification results qualitatively in Figure.~\ref{fig:sort}. The preliminary results qualitatively validate our proposed system, and further research will work on improving each components and perform user study to prove the efficiency and effectiveness of the system, and also on how to model the dynamic nature of the user interests.

{\small
    \bibliographystyle{abbrv}
    \bibliography{refers}

\begin{thebibliography}{10}

\bibitem{frey07affinitypropagation}
B.~J. Frey and D.~Dueck.
\newblock Clustering by passing messages between data points.
\newblock {\em Science}, 315:972--976, 2007.

\bibitem{Geng:2014:OKU:2647868.2654950}
X.~Geng, H.~Zhang, Z.~Song, Y.~Yang, H.~Luan, and T.-S. Chua.
\newblock One of a kind: User profiling by social curation.
\newblock MM'14.

\bibitem{Kimura:2013:ICD:2502081.2502149}
A.~Kimura, K.~Ishiguro, M.~Yamada, A.~Marcos~Alvarez, K.~Kataoka, and
  K.~Murasaki.
\newblock Image context discovery from socially curated contents.
\newblock MM '13.

\bibitem{krizhevsky2012imagenet}
A.~Krizhevsky, I.~Sutskever, and G.~E. Hinton.
\newblock Imagenet classification with deep convolutional neural networks.
\newblock NIPS '12.

\bibitem{pennebaker2001linguistic}
J.~Pennebaker, M.~Francis, and R.~Booth.
\newblock {\em Linguistic inquiry and word count [computer software]}.
\newblock Mahwah, NJ: Erlbaum Publishers, 2001.

\bibitem{read2010scalable}
J.~Read.
\newblock Scalable multi-label classification.
\newblock 2010.

\bibitem{Ren:2014:HMC:2600428.2609595}
Z.~Ren, M.-H. Peetz, S.~Liang, W.~van Dolen, and M.~de~Rijke.
\newblock Hierarchical multi-label classification of social text streams.
\newblock SIGIR '14. ACM.

\bibitem{9783540749578}
G.~Tsoumakas and I.~Vlahavas.
\newblock Random k-labelsets: An ensemble method for multilabel classification.
\newblock In {\em Machine Learning: ECML 2007}. Springer Berlin Heidelberg,
  2007.

\bibitem{Weinberger:2009:FHL:1553374.1553516}
K.~Weinberger, A.~Dasgupta, J.~Langford, A.~Smola, and J.~Attenberg.
\newblock Feature hashing for large scale multitask learning.
\newblock ICML '09, New York, NY, USA. ACM.

\bibitem{Yan:2014:MCA:2647868.2654920}
M.~Yan, J.~Sang, and C.~Xu.
\newblock Mining cross-network association for youtube video promotion.
\newblock MM '14.

\end{thebibliography}
}

\end{document}